%% file: main.tex
\documentclass{article}
\PassOptionsToPackage{numbers,sort&compress}{natbib}
\usepackage[preprint]{neurips_2026}
\usepackage[utf8]{inputenc}
\usepackage[T1]{fontenc}
\usepackage{graphicx}
\usepackage{booktabs}
\usepackage{placeins}
\usepackage{tabularx}
\usepackage{array}
\usepackage{longtable}
\usepackage{amsmath}
\usepackage{xcolor}
\usepackage{microtype}
\usepackage{hyperref}
\hypersetup{
  hidelinks,
  pdftitle={A Description-Derived DDI Benchmark Reveals Split-Dependent Molecular Representation Rankings},
  pdfauthor={}
}
\graphicspath{{figures/}}
\input{rps_assets/macros}

\title{A Description-Derived Drug-Drug Interaction Benchmark Reveals Split-Dependent Molecular Representation Rankings}
\workshoptitle{Representations for the Physical Sciences}

\author{
  Manel Gil-Sorribes$^{1}$ \quad
  Isaac Filella-Merce$^{1}$ \quad
  Alexis Molina$^{1,2}$ \\[0.5em]
  $^{1}$Nostrum Biodiscovery \\
  $^{2}$AITHYRA \\[0.5em]
  \texttt{manel.gil@nostrumbiodiscovery.com} \\
  \texttt{alexis.molina@nostrumbiodiscovery.com}
}

\makeatletter
\renewcommand{\@noticestring}{Submitted to the 40th Conference on Neural Information Processing Systems (NeurIPS 2026). Workshop: Agentic Systems for Molecular Sciences. Do not distribute.}
\makeatother

\begin{document}
\maketitle

\begin{abstract}
\input{abstract}
\end{abstract}

\section{Introduction}

Drug-drug interaction (DDI) prediction differs from many molecular property
tasks because the prediction target belongs to a pair rather than to an
individual molecule. A downstream predictor must combine two molecular
representations to infer an interaction that neither single-molecule
representation encodes directly. DrugBank provides curator-authored natural-language interaction
descriptions at scale \cite{Knox2024DrugBank6}. However, turning them into categorical prediction targets is itself a
dataset-construction choice.

Evaluation introduces a second choice. Existing DDI studies use pair- and
drug-holdout protocols \cite{Zhong2024MeTDDI}, as well as knowledge-graph
formulations \cite{Yu2021SumGNN}. Holding out pairs does not necessarily hold
out their molecular content: test pairs may still contain training-seen drugs,
scaffolds, or closely related structures. Scaffold-aware evaluation addresses
this issue \cite{Wu2018MoleculeNet,Bemis1996Murcko}.

Morgan fingerprints encode fixed local environments
\cite{Rogers2010ECFP}, MoLFormer and ChemBERTa learn molecular-string
representations \cite{Ross2022MoLFormer,Chithrananda2020ChemBERTa}, and
graph-derived representations encode molecular connectivity through graph
neural networks and message passing
\cite{Kipf2017GCN,Gilmer2017NeuralMessagePassing}. Previous DDI
experiments already suggested that fingerprints can be favored when substantial
chemical overlap remains \cite{GilSorribes2026OvercomplexityDDI}. We therefore
ask whether a representation ranking persists when the labels are fixed but the
molecular novelty of the test population changes.

To evaluate the effects of distinct molecular representations in DDI across
different holdout regimes, we construct a dataset of \PairCount{} unordered DrugBank pairs with
\CategoryCount{} text-derived interaction categories. We use dataset to
refer to these labeled pairs, and benchmark to refer to the dataset
together with the fixed split manifests and evaluation protocol. The benchmark
contains random pair, one-drug-unseen, two-drug-unseen, and scaffold-disjoint
regimes, allowing the same representation grid to be evaluated as overlap with
training is restricted.

We summarize our contributions as:
\begin{itemize}
    \item a DDI dataset containing \PairCount{} unordered DrugBank pairs and
    \CategoryCount{} traceable, description-derived categories,

    \item a benchmark with four evaluation regimes that explicitly control
    pair, drug, scaffold, and molecular-similarity overlap, and

    \item experiments repeated across \FiveSeeds{} model seeds showing that
    representation rankings change with test-set novelty, while classifier-head
    complexity has a smaller effect on the conclusions.
\end{itemize}

\section{Dataset construction and benchmark design}

\subsection{DDI dataset and text-derived labels}

Starting from DrugBank DDI records, directed entries are
canonicalized to one row per unordered DrugBank ID pair, producing 1,392,791 labeled pairs across 4,437 drugs, with representative
description examples provided in Appendix~\ref{app:clustering}, Table \ref{tab:representative-patterns}. The subject and affected drug
names are then removed from each description to reduce direct identity cues.
The cleaned descriptions are embedded with a biomedical
SentenceTransformer using a PubMedBERT backbone
\cite{Gu2021PubMedBERT,Reimers2019SentenceBERT}.
Table~\ref{tab:dataset-subsets} summarizes molecular-structure coverage in the
labeled dataset. The reported split manifests are constructed from the full
labeled pair table, with unavailable molecular features handled during feature
construction. Additional dataset construction details are provided in Appendix~\ref{app:dataset}.

\begin{table*}[htbp]
\centering
\scriptsize
\renewcommand{\arraystretch}{1.15}
\caption{Dataset and molecular-structure coverage. The full labeled dataset
provides the pair population used to construct the reported split manifests.
The structure-resolved subset contains pairs for which both endpoints have
valid RDKit-parsable SMILES.}
\begin{tabularx}{\textwidth}{>{\raggedright\arraybackslash}p{0.30\textwidth}rr>{\raggedright\arraybackslash}p{0.05\textwidth}>{\raggedright\arraybackslash}X}
\toprule
Table or subset & Rows & Pairs & Drugs & Role \\
\midrule
Full labeled DDI dataset & 1,392,791 & 1,392,791 & 4,437 & Base for reported split manifests \\
Pairs with valid SMILES for both drugs & 1,158,783 & 1,158,783 & 3,468 & Fully structure-resolved subset \\
\bottomrule
\label{tab:dataset-subsets}
\end{tabularx}
\end{table*}

To derive categorical labels for each drug pair, we clustered the corresponding
DDI description embeddings. We first sampled 50,000 descriptions to
evaluate alternative clustering configurations. Among the tested agglomerative settings, the six-category solution achieved the
strongest combined separation-and-balance score. MiniBatchKMeans was then used
for scalable assignment of the complete dataset \cite{Sculley2010WebScaleKMeans}. Table~\ref{tab:label-summary}
summarizes the recurring description signals associated with the resulting
categories. Full clustering diagnostics and representative wording patterns
are provided in Appendix~\ref{app:clustering}.

\begin{table}[h]
\centering
\small
\caption{Recurring DrugBank description signals represented by the six
text-derived categories. The categories summarize description language
rather than expert-adjudicated mechanisms.}
\label{tab:label-summary}
\begin{tabularx}{\linewidth}{@{}cX@{}}
\toprule
Category & Recurring description signals \\
\midrule
0 & exposure, serum concentration, protein binding, metabolites \\
1 & exposure, efficacy, adverse effects, bioavailability \\
2 & excretion, serum level, efficacy changes \\
3 & CNS effects, blood pressure, coagulation, cardiac effects \\
4 & bleeding, QTc, hypertension, electrolyte and safety effects \\
5 & metabolism and absorption \\
\bottomrule
\end{tabularx}
\end{table}

\begin{figure}[h]
  \centering
  \includegraphics[width=0.94\linewidth]{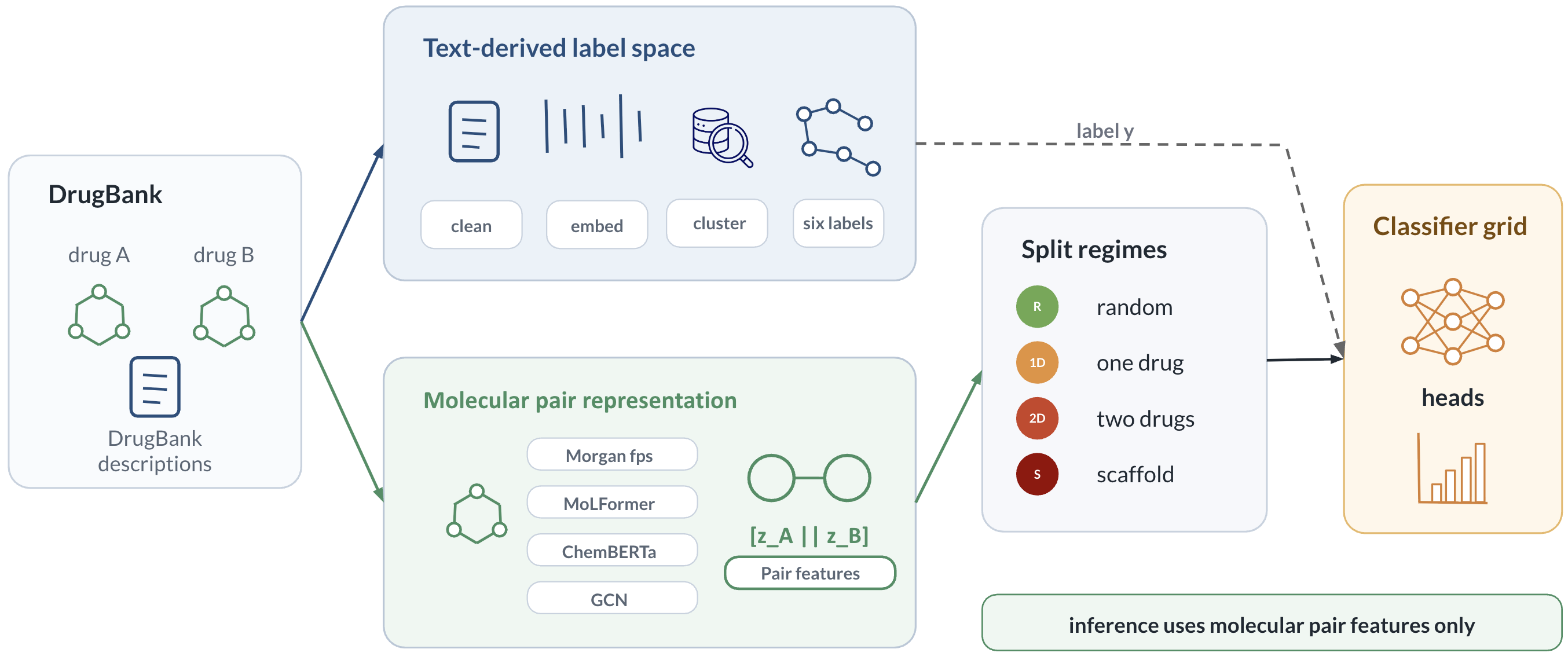}
  \caption{Benchmark construction and evaluation. DrugBank descriptions are used to obtain
  the six text-derived labels, while molecular representations are the
  inference inputs evaluated under pair-, drug-, and scaffold-level
  holdouts.}
  \label{fig:rps-workflow}
\end{figure}

Because the categories remain traceable to recurring DrugBank wording,
their content can be inspected against the source descriptions. The text encoder is used to derive the benchmark labels and is not evaluated
as a molecular representation for DDI prediction (Figure~\ref{fig:rps-workflow}).

We construct the DDI benchmark by combining this fixed label space with four predefined train, validation, and test splits.
The labels therefore remain unchanged across evaluation regimes, while the
amount and type of molecular information shared between training and test are
controlled. This separation is central to the benchmark: differences between
splits can be interpreted against the same prediction target rather than
against independently constructed DDI datasets.

\subsection{Benchmark regimes and representation evaluation}

Four evaluation regimes alter the held-out population while keeping the
label space and training protocol fixed. Random pair holdout evaluates unseen
combinations while retaining high overlap in the individual drugs.
One-drug-unseen evaluation requires exactly one test endpoint to be absent from
training, whereas two-drug-unseen evaluation removes both. Scaffold-disjoint
evaluation additionally enforces zero train-test DrugBank ID,
canonical SMILES, and Bemis-Murcko-framework overlap
\cite{Bemis1996Murcko}. ECFP4 similarities \cite{Rogers2010ECFP} are computed
with RDKit \cite{Landrum2025RDKit}, and the maximum validation/test-to-train
Tanimoto similarity is constrained to \ScaffoldTanimoto{}. Table~\ref{tab:split-sizes-main} reports the resulting split sizes.
All regimes have zero train-test unordered-pair overlap. Scaffold-disjoint
evaluation additionally has zero train-test drug-identity overlap and,
among structure-resolved compounds, zero canonical SMILES and
Bemis-Murcko-scaffold overlap, with maximum ECFP4 Tanimoto similarity
of 0.70.

\begin{table}[h]
\centering
\scriptsize
\caption{Train, validation, and test sizes across benchmark regimes. Drug-
and scaffold-holdout settings retain only pairs satisfying their corresponding
partition constraints.}
\label{tab:split-sizes-main}
\begin{tabular}{lrrrr}
\toprule
Split & Train & Val. & Test & Total \\
\midrule
Random & 974,954 & 139,279 & 278,558 & 1,392,791 \\
One drug unseen & 675,377 & 191,699 & 397,605 & 1,264,681 \\
Two drugs unseen & 691,248 & 13,179 & 54,956 & 759,383 \\
Scaffold-disjoint & 806,341 & 10,970 & 36,621 & 853,932 \\
\bottomrule
\end{tabular}
\end{table}

All scaffold-disjoint validation and test pairs are structure-resolved.
613,853 of the 806,341 training pairs have valid structures for both endpoints,
with unavailable molecular features handled by the zero-vector fallback
described in Appendix \ref{app:models}.

The benchmark evaluates four forms of molecular input that differ in both
construction and pretraining signal. Morgan fingerprints are fixed,
rule-defined circular descriptors \cite{Rogers2010ECFP}. MoLFormer and ChemBERTa use self-supervised molecular-string
representations learned with masked-token objectives
\cite{Ross2022MoLFormer,Chithrananda2020ChemBERTa}. GCN features use the Tensor-DTI drug encoder trained on PCBA-1328 \cite{GilSorribes2026TensorDTI,Kim2023PubChem}. Pair features are concatenated in canonical DrugBank ID order. All four representations are evaluated with the same predefined
classifier-head grid and five model seeds to test sensitivity to downstream
classifier capacity. Architectural and optimization
details are given in Appendix~\ref{app:models}.

\section{What the benchmark reveals}

The benchmark reveals a clear representation-ranking reversal as molecular
novelty increases: Morgan fingerprints perform best in the high-overlap random
split, whereas Tensor-DTI graph-derived features lead the one-drug-unseen,
two-drug-unseen, and scaffold-disjoint regimes.

\input{rps_assets/table_main_results}

\begin{figure}[h]
    \centering
    \includegraphics[width=\linewidth]{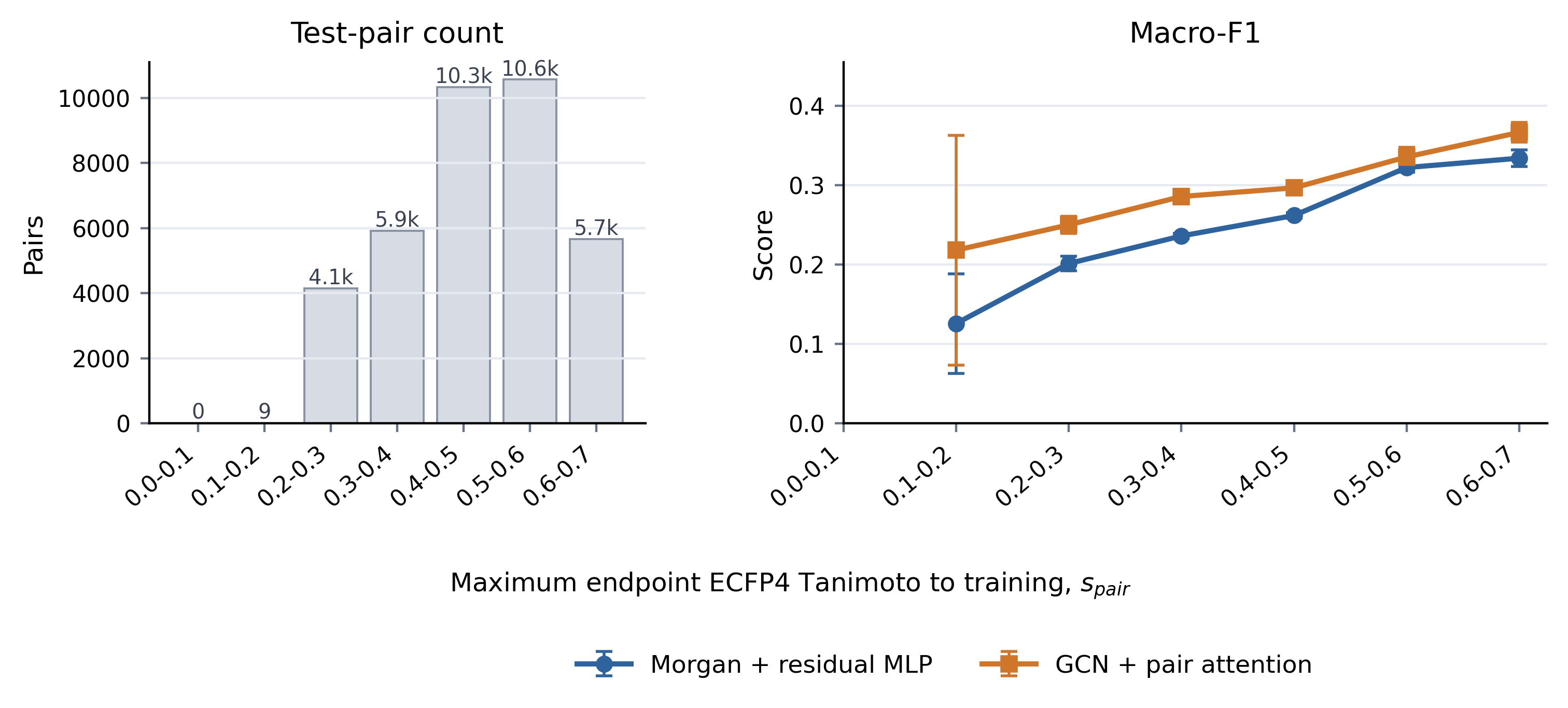}
    \caption{Scaffold-disjoint performance as a function of similarity to the
training chemistry. Left: test-pair counts by maximum endpoint-to-training
ECFP4 Tanimoto similarity. Right: five-seed macro-F1 for Morgan and GCN.
Error bars show mean $\pm$ standard deviation.}
    \label{fig:scaffold-tanimoto-bins}
\end{figure}

Table~\ref{tab:main_splits_results} quantifies this reversal. The GCN margin
over Morgan increases from 0.004 macro-F1 in the one-drug-unseen regime to
0.020 in the two-drug-unseen regime and 0.037 in the scaffold-disjoint regime.
Its margin over the strongest non-GCN alternative also increases across these
settings, from 0.003 to 0.012 and 0.016, respectively. Within the scaffold-disjoint test set, GCN also outperforms Morgan across
the populated similarity bins (Figure~\ref{fig:scaffold-tanimoto-bins}). A benchmark restricted
to the random split would therefore rank Morgan fingerprints first and obscure
the reversal observed under molecular novelty. Because the prediction target,
representation grid, and training protocol are held fixed across regimes, the
results show that representation rankings depend strongly on the degree of
train-test molecular overlap. Additional configuration rankings, paired
comparisons, and diagnostic metrics are provided in
Appendix~\ref{app:extended-results}.

Classifier-head variation is smaller than the representation and split effects,
and changing the head does not alter the winning representation within any
regime.

\section{Limitations}

The six categories are derived from recurring DrugBank description language
rather than expert-adjudicated mechanisms, and may inherit annotation or
wording biases from the source. The molecular representations also differ in
architecture and pretraining data, so the observed graph-derived
advantage cannot be attributed to any one of these factors in isolation.

\section{Conclusion}

Representation rankings in DDI prediction change substantially with the evaluation
regime. We also introduce a description-derived DDI benchmark that combines a fixed
six-category label space with pair-, drug-, and scaffold-level holdouts and
explicit molecular-similarity controls. Across five seeds, Morgan fingerprints lead under high-overlap random
sampling, whereas GCN features lead the one-drug-unseen, two-drug-unseen, and
scaffold-disjoint settings. Classifier-head changes affect absolute performance
within a representation but do not alter this split-dependent ranking in the
evaluated grid.

This work shows that the choice of representation depends on
what information is allowed to recur between training and test. Random pair
splits primarily evaluate unseen combinations while drug identities and local
chemical patterns remain well represented in training. Drug-holdout and
scaffold-disjoint settings instead test complementary forms of transfer
to unseen compounds and chemical neighborhoods. No single regime is intrinsically correct for every application, but the
evaluation split should reflect the intended deployment setting.

These results reinforce the need to compare learned molecular representations
against simple fingerprint baselines under explicit overlap controls. Reporting
only a random-split score can favor a representation that exploits recurring
chemical information while obscuring a different ranking under molecular novelty.
Our benchmark provides one way to make these conditions explicit, and DDI
benchmarks more generally should report performance together with the degree of drug, scaffold, and molecular-similarity overlap between training and test. Without
this context, representation rankings may reflect the evaluation design as much
as the transferable information captured by the molecular features.

\clearpage
\bibliographystyle{unsrtnat}
\bibliography{rps_assets/bibliography}
\input{rps_assets/appendix}
\end{document}

%% file: rps_assets/macros.tex
\newcommand{\PairCount}{1,392,791}
\newcommand{\CategoryCount}{six}
\newcommand{\FiveSeeds}{five}

\newcommand{\ScaffoldTanimoto}{0.70}

%% file: abstract.tex
Drug-drug interaction (DDI) prediction is clinically important and poses a
distinct challenge for molecular representation learning. Unlike
single-molecule property prediction, DDI requires a predictor to compose two
molecular representations into a relational outcome. DrugBank provides curator-authored DDI descriptions rather than ready-made categorical labels, so a categorical prediction target must first be derived from the text. We construct a dataset of DDI pairs by mapping processed descriptions to six text-derived categories and define a benchmark with four holdout regimes that restrict train-test
chemical overlap from random pairs to scaffold-disjoint evaluation. We then use this benchmark to compare molecular fingerprints, pretrained sequence representations, and graph-derived
features across five seeds and multiple classifier heads. Molecular
fingerprints lead the high-overlap random split, whereas graph-derived
representations lead the drug- and scaffold-holdout regimes, with the gap
increasing as molecular overlap is restricted. Changing the classifier head
affects absolute performance but not the relative representation ranking.
These results show that representation rankings change with test-set novelty
under a fixed description-derived label space.

%% file: rps_assets/table_main_results.tex
\begin{table*}[h]
\centering
\small
\caption{Representation performance across split regimes. Each cell reports the best observed mean macro-F1 $\pm$ standard deviation across five seeds within the predefined classifier-head grid. Bold marks the leading representation within each regime, and underlining marks the second-best.}
\label{tab:main_splits_results}
\setlength{\tabcolsep}{4pt}
\begin{tabular}{lcccc}
\toprule
Representation & Random & One drug unseen & Two drugs unseen & Scaffold-disjoint \\
\midrule
Morgan pair & \textbf{0.899 $\pm$ 0.000} & 0.631 $\pm$ 0.002 & 0.412 $\pm$ 0.003 & 0.287 $\pm$ 0.005 \\
MoLFormer pair & \underline{0.876 $\pm$ 0.001} & \underline{0.632 $\pm$ 0.001} & \underline{0.420 $\pm$ 0.005} & \underline{0.308 $\pm$ 0.005} \\
ChemBERTa pair & 0.866 $\pm$ 0.001 & 0.623 $\pm$ 0.001 & 0.398 $\pm$ 0.005 & 0.299 $\pm$ 0.005 \\
GCN pair & 0.856 $\pm$ 0.001 & \textbf{0.635 $\pm$ 0.001} & \textbf{0.432 $\pm$ 0.004} & \textbf{0.324 $\pm$ 0.005} \\
\bottomrule
\end{tabular}
\end{table*}

%% file: rps_assets/appendix.tex
\clearpage
\appendix

\FloatBarrier
\section{Additional dataset construction details}
\label{app:dataset}

DrugBank interaction records are canonicalized to one row per unordered
DrugBank ID pair. Before text embedding, the subject and affected drug names
are removed from each interaction description to reduce direct identity cues.
The cleaned corpus contains 417 normalized description strings, of which 380
occur in the evaluated dataset.

Among the full labeled pairs, 1,158,783 have valid RDKit-parsable SMILES for
both endpoints, corresponding to 3,468 drugs. The reported split manifests,
however, are constructed from the full labeled pair table. Pairs containing
drugs absent from a molecular feature bank are retained and the unavailable
endpoint representation is replaced by a zero vector during feature
construction. DrugBank source data remain subject to the applicable
license. Benchmark scripts, aggregate artifacts, and manuscript resources will
be released separately.

\FloatBarrier
\section{Text clustering and label interpretation}
\label{app:clustering}
Cleaned descriptions are embedded with the SentenceTransformer checkpoint
\texttt{pritamdeka/S-PubMedBert-MS-MARCO}\footnote{
\url{https://huggingface.co/pritamdeka/S-PubMedBert-MS-MARCO}},
which uses a PubMedBERT backbone within the Sentence-BERT framework
\cite{Gu2021PubMedBERT,Reimers2019SentenceBERT}. Candidate cluster counts are evaluated on a sample of 50,000 description embeddings before scalable assignment over the full table. The combined selection score is
\[
\mathrm{selection\ score}=\mathrm{silhouette}-\mathrm{Davies\text{-}Bouldin}+\mathrm{cluster\ size\ balance},
\]
where cluster-size balance is the minimum cluster size divided by the maximum cluster size. Calinski-Harabasz is reported separately as an additional diagnostic and is not included in the combined score.

Table~\ref{tab:cluster-diagnostics} compares the sampled candidate granularities. Agglomerative clustering with \(k=6\) has the strongest listed combined score. MiniBatchKMeans \cite{Sculley2010WebScaleKMeans} is then used to assign the full corpus at the selected six-category granularity, which is a practical benchmark choice rather than a mathematically unique optimum.

\input{rps_assets/tables/table_cluster_selection}

\begin{figure}[ht]
  \centering
  \includegraphics[width=0.78\linewidth]{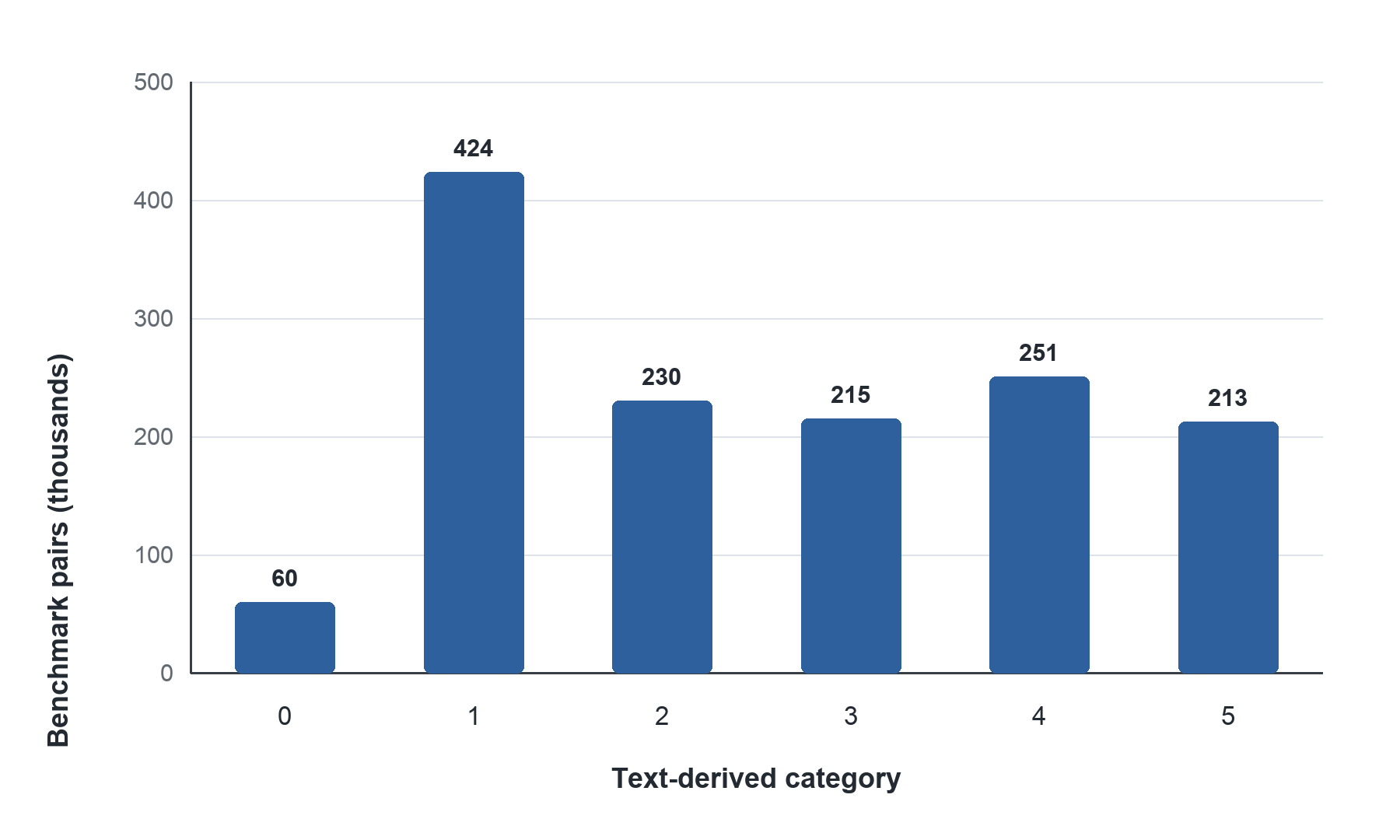}
  \caption{Cluster-size distribution for the six text-derived interaction-description categories.}
  \label{fig:cluster-sizes}
\end{figure}

Benchmark prevalence is frequency-weighted because some cleaned templates recur many times. Figure~\ref{fig:cluster-sizes} and Table~\ref{tab:cluster-statistics} therefore report both pair-weighted prevalence and unique-template counts.

\input{rps_assets/tables/table_cluster_template_stats}

Clusters 0, 1, 2, and 5 are dominated by a small number of recurring templates, whereas clusters 3 and 4 contain greater template diversity. Reporting both pair-weighted counts and unique-template counts makes this difference visible.

Tables~\ref{tab:cluster-summaries} and~\ref{tab:representative-patterns} provide description-level interpretability aids. They summarize recurring wording patterns used by the benchmark target.
They are not an external biological validation.

\input{rps_assets/tables/table_cluster_biology}
\input{rps_assets/tables/table_cluster_examples}

The categories can be traced to recurring wording in the source descriptions,
but should not be interpreted as independently adjudicated mechanistic or
clinical labels.

\FloatBarrier
\section{Molecular representations and model heads}
\label{app:models}
This section separates molecular representation choice, pair construction, classifier-head capacity, and the common training protocol. Morgan fingerprints use radius 2 and 2048 bits, MoLFormer uses \texttt{ibm/MoLFormer-XL-both-10pct}, ChemBERTa uses \texttt{DeepChem/ChemBERTa-77M-MLM}, and GCN features use the Tensor-DTI drug encoder trained on PCBA-1328 \cite{GilSorribes2026TensorDTI,Kim2023PubChem}. All pair features concatenate endpoint vectors in canonical DrugBank ID order.

Table~\ref{tab:model-hyperparameters} gives the model-head and optimization settings. The auxiliary-regression head uses a description embedding only as a
training-time target. Interaction text is not an inference-time model input. Its purpose is to regularize the molecular pair representation toward the text-derived label space.

\input{rps_assets/tables/table_hyperparameters}

The shared representation and optimization controls help attribute performance differences to the representation, split regime, and head choices recorded in the result tables. Data construction uses seed 42, and model results are aggregated over seeds 11, 22, 33, 44, and 55. Endpoint vectors use canonical DrugBank ID order, and feature standardization is fit on the training partition only.

\FloatBarrier
\section{Split construction and validation}
\label{app:splits}
The four regimes isolate complementary forms of generalization. Random evaluation tests unseen pair combinations while retaining high
individual-drug overlap. One-drug-unseen and two-drug-unseen evaluation
require one or both held-out endpoints, respectively. Scaffold-disjoint
evaluation restricts drug identity, canonical SMILES, Bemis-Murcko scaffold,
and molecular-similarity overlap. These regimes should therefore be read as complementary generalization tests rather than as a single strictly monotonic difficulty scale.

The split sizes are reported in Table \ref{tab:split-sizes-main}. Table \ref{tab:split-validation} provides the corresponding
overlap and assignment checks.

All 47,591 scaffold-disjoint validation and test pairs have valid molecular
structures for both endpoints. Of the 806,341 training pairs, 613,853 have
two valid structures, while 192,488 contain at least one unavailable or invalid
structure and are retained using the zero-vector feature fallback described in
Appendix \ref{app:models}.

One-drug-unseen has the largest test set because many pairs satisfy its one-held-out-endpoint rule. The smaller scaffold-disjoint validation and test sets follow from the globally consistent molecular partition and cross-partition exclusions, rather than from arbitrary subsampling. Table~\ref{tab:split-validation} records the overlap and assignment checks.

\input{rps_assets/tables/table_split_leakage_audit}

Every regime has zero train-test unordered-pair overlap. The scaffold-disjoint split additionally has zero train-test DrugBank ID
overlap and, among structure-resolved compounds, zero canonical SMILES and
Bemis-Murcko scaffold overlap. The molecular-similarity constraint is verified in
Table~\ref{tab:split-validation}. These controls test chemical-neighborhood generalization on a different axis from the endpoint novelty isolated by two-drug-unseen evaluation.

\section{Extended results}
\label{app:extended-results}
The following subsections retain the configuration-level evidence behind the main-paper winner statements. They distinguish representation effects from classifier-head variation and report complementary probability-based metrics without changing macro-F1 as the primary ranking metric.

\subsection{Top-performing configurations}
Table~\ref{tab:top-configurations} reports the top three five-seed Torch configurations within each split, selecting representation and head separately for each regime. It supports the winner statements in the main text.

\input{rps_assets/tables/table_top_configs}

Morgan occupies the top random positions, whereas GCN leads one-drug-unseen, two-drug-unseen, and scaffold-disjoint evaluation. Pair attention gives the highest observed scaffold-disjoint GCN result.

\subsection{Paired representation and head comparisons}
Differences in Table~\ref{tab:paired-comparisons} are computed seed-wise with the same five model seeds, so the intervals describe paired configuration margins rather than independent-run comparisons.

\input{rps_assets/tables/table_statistical_comparisons}

The GCN-MoLFormer margin is small but positive for one-drug-unseen evaluation and larger for two-drug-unseen evaluation. The listed GCN configuration exceeds the listed MoLFormer configuration in scaffold-disjoint evaluation because its paired interval excludes zero. By contrast, the pair-attention versus residual-GCN interval includes zero, so head superiority is not conclusive.

\subsection{Probability-based diagnostics}
Alongside macro-F1, Table~\ref{tab:probability-diagnostics} reports balanced accuracy, MCC, macro AUROC, macro AUPRC, and confidence ECE \cite{Guo2017Calibration} for the leading configuration in each regime. These metrics assess complementary aspects of class-balanced discrimination
and confidence calibration. ECE is not a full epistemic-uncertainty analysis.

\input{rps_assets/tables/table_probability_metrics}

Discrimination and calibration both degrade under stricter generalization. The strong random result is not preserved under unseen-drug and scaffold-disjoint evaluation, while macro-F1 remains the primary metric for comparing the imbalanced six-category task.

\subsection{Comparison of generalization regimes}
Each row in Table~\ref{tab:split-comparison} compares the same training setup, representation, and head across two-drug-unseen and scaffold-disjoint evaluation. Because the evaluated pair populations differ, the comparison is descriptive rather than a paired test on identical examples.

\input{rps_assets/tables/table_split_difficulty_comparison}

Scaffold-disjoint macro-F1 is lower in all 36 matched comparisons. This supports treating it as a distinct chemical-generalization stress test rather than another drug-holdout split.

\FloatBarrier
\subsection{Classifier-head summary}
Table~\ref{tab:head-summary} averages over representations and therefore provides a supporting view of head robustness rather than a replacement for representation-specific comparisons.

\input{rps_assets/tables/table_head_summary}

Residual MLP is the most consistently strong average head. This does not contradict pair attention producing the single best scaffold-disjoint GCN configuration, because head-level averages should not replace representation-specific comparisons.

%% file: rps_assets/tables/table_cluster_selection.tex
\begin{table*}[htbp]
\centering
\small
\caption{Cluster-granularity diagnostics on a sample of 50,000 SentenceTransformer description embeddings. The combined selection score is silhouette minus Davies-Bouldin plus cluster-size balance. Calinski-Harabasz is reported but not included in the score. Agglomerative clustering with \(k=6\) has the strongest listed score and motivates the six-category label granularity used for full-corpus assignment.}
\label{tab:cluster-diagnostics}
\setlength{\tabcolsep}{4pt}
\begin{tabular}{llrrrrr}
\toprule
Method & k & Sil. & Davies-Bouldin & Calinski-Harabasz & Balance & Score \\
\midrule
Agglomerative & 6 & 0.565 & 1.205 & 19,135 & 0.719 & 0.079 \\
Agglomerative & 8 & 0.644 & 1.045 & 20,807 & 0.253 & -0.148 \\
Agglomerative & 50 & 0.943 & 1.095 & 34,605 & 0.003 & -0.148 \\
Agglomerative & 30 & 0.892 & 1.112 & 24,878 & 0.020 & -0.199 \\
Agglomerative & 5 & 0.506 & 1.184 & 18,390 & 0.386 & -0.292 \\
\bottomrule
\end{tabular}
\end{table*}

%% file: rps_assets/tables/table_cluster_template_stats.tex
\begin{table}[htbp]
\centering
\scriptsize
\renewcommand{\arraystretch}{1.12}
\caption{Cluster statistics for the benchmark table used in model evaluation. Pair percentages are frequency-weighted by benchmark rows, whereas template counts treat normalized descriptions as unique strings.}
\label{tab:cluster-statistics}
\begin{tabular}{rrrrr}
\toprule
Cluster & Pairs & Pair \% & Cleaned templates & Top-5 template row \% \\
\midrule
0 & 60,169 & 4.3 & 23 & 99.8 \\
1 & 423,895 & 30.4 & 37 & 99.8 \\
2 & 230,493 & 16.5 & 17 & 99.9 \\
3 & 215,150 & 15.4 & 104 & 69.3 \\
4 & 250,567 & 18.0 & 182 & 54.7 \\
5 & 212,517 & 15.3 & 18 & 100.0 \\
\bottomrule
\end{tabular}
\end{table}

%% file: rps_assets/tables/table_cluster_biology.tex
\begin{table*}[t]
\centering
\scriptsize
\renewcommand{\arraystretch}{1.15}
\caption{Textual summary of the six text-derived interaction-description categories. The summaries link each benchmark label to recurring DrugBank wording patterns and provide a qualitative view of label interpretability.}
\label{tab:cluster-summaries}
\begin{tabularx}{\textwidth}{r X X}
\toprule
Cluster & Description-level summary & Representative signals \\
\midrule
0 & pharmacokinetic exposure changes, metabolic inhibition or induction, protein binding and distribution, therapeutic effect modulation & serum, concentration, protein, binding, active, metabolites, metabolite \\
1 & pharmacokinetic exposure changes, pharmacodynamic toxicity additivity, therapeutic effect modulation & effects, adverse, risk, severity, therapeutic, efficacy, bioavailability \\
2 & pharmacokinetic exposure changes, therapeutic effect modulation & level, excretion, serum, efficacy, reduction \\
3 & therapeutic effect modulation, cardiac electrophysiology, central nervous system depression, blood pressure and vascular tone, coagulation and bleeding, glucose regulation & depressant, activities, nervous, central, cns, antihypertensive, hypotensive, arrhythmogenic \\
4 & pharmacodynamic toxicity additivity, cardiac electrophysiology, blood pressure and vascular tone, coagulation and bleeding, renal and electrolyte safety, glucose regulation & severity, risk, hypertension, bleeding, prolongation, qtc, hyperkalemia, hemorrhage \\
5 & pharmacokinetic exposure changes, metabolic inhibition or induction, therapeutic effect modulation & metabolism, absorption \\
\bottomrule
\end{tabularx}
\end{table*}

%% file: rps_assets/tables/table_cluster_examples.tex
\begin{table*}[htbp]
\centering
\scriptsize
\renewcommand{\arraystretch}{1.15}
\caption{Representative wording patterns for the text-derived labels. These template-like patterns are extracted from frequent cleaned descriptions with drug names omitted and show recurring interaction-description language.}
\label{tab:representative-patterns}
\begin{tabularx}{\textwidth}{r X}
\toprule
Cluster & Representative description patterns \\
\midrule
0 & serum concentration can be increased when combined, protein binding and active-metabolite language \\
1 & therapeutic efficacy can be decreased when used in combination, therapeutic efficacy can be increased when used in combination \\
2 & excretion rate can be decreased, resulting in higher serum level, serum-level and excretion language \\
3 & anticoagulant activities can be increased, central nervous system depressant activities can be increased \\
4 & risk or severity of bleeding can be increased, risk or severity of hypertension can be increased \\
5 & metabolism can be increased when combined, metabolism can be decreased when combined \\
\bottomrule
\end{tabularx}
\end{table*}

%% file: rps_assets/tables/table_hyperparameters.tex
\begin{table*}[htbp]
\centering
\small
\renewcommand{\arraystretch}{1.12}
\caption{Model-head and training hyperparameters used in the final streaming Torch experiments. Values are taken from the training script used for the benchmark grid.}
\label{tab:model-hyperparameters}
\begin{tabularx}{\textwidth}{l>{\raggedright\arraybackslash}X}
\toprule
Component & Setting \\
\midrule
Linear head & Single linear layer from input features to six classes \\
MLP head & 512-unit hidden layer, ReLU, dropout 0.2, 256-unit hidden layer, ReLU, dropout 0.2, six-class output \\
Residual MLP & 512-dimensional projection followed by two residual blocks, each block uses linear-ReLU-dropout 0.2-linear with layer normalization \\
Gated MLP & 512-dimensional value branch multiplied by a 512-dimensional sigmoid gate, followed by dropout 0.2, 256-unit hidden layer, dropout 0.2, and output layer \\
Feature-attention MLP & 512-dimensional value branch multiplied by a learned sigmoid feature-attention vector, followed by the same 256-unit head as the gated MLP \\
Pair-attention head & Two drug vectors projected to 256 dimensions, four-head self-attention with dropout 0.1 over the two tokens, layer-normalized 512-dimensional flattened token output passed to a 512-unit head with dropout 0.2 \\
Auxiliary-regression head & 512-unit and 256-unit encoder with dropout 0.2, cross-entropy classifier plus 768-dimensional regression head trained with cosine embedding loss, \(\lambda_{\mathrm{cls}}=1.0\), \(\lambda_{\mathrm{reg}}=1.0\) \\
Optimization & AdamW, learning rate \(10^{-3}\), weight decay \(10^{-4}\), 30 epochs, batch size 1024, evaluation batch size 8192 \\
Data handling & Streaming feature construction, train-set feature standardization applied to held-out batches, final Torch runs use all rows assigned to each split, the extended comparison additionally includes eight shallow baseline configurations. Drugs absent from a representation feature bank are assigned zero vectors. \\
Seeds & Data construction seed 42, model seeds 11, 22, 33, 44, and 55 for the Torch grid \\
\bottomrule
\end{tabularx}
\end{table*}

%% file: rps_assets/tables/table_split_leakage_audit.tex
\begin{table*}[h]
\centering
\scriptsize
\caption{Split-validation checks for the split suite. Train-test pair overlap is zero in every setting. The scaffold-disjoint split additionally has zero train-test DrugBank ID, canonical SMILES, and Bemis-Murcko scaffold overlap. Its maximum validation/test-to-train ECFP4 Tanimoto similarity is 0.70, with no value above the threshold.}
\label{tab:split-validation}
\begin{tabularx}{\textwidth}{lrrX}
\toprule
Split & Pair overlap & Train-test drug overlap & Validation check \\
\midrule
Random & 0 & 4,169/4,207 & row-level shuffle, most test drugs remain training-seen \\
One drug unseen & 0 & 2,995/3,874 & 0 rows violate the one-unseen-drug rule \\
Two drugs unseen & 0 & 0/862 & 0 rows violate the two-unseen-drug rule \\
Scaffold-disjoint & 0 & 0/656 & 0 canonical SMILES/scaffold overlap, maximum ECFP4 Tanimoto 0.70 \\
\bottomrule
\end{tabularx}
\end{table*}

%% file: rps_assets/tables/table_top_configs.tex
\begin{table*}[htbp]
\centering
\small
\caption{Top three five-seed Torch configurations per evaluated split. Morgan pair features lead random evaluation, whereas GCN pair features lead all three drug- and scaffold-holdout settings. The scaffold-disjoint setting is also the only leading setting in which pair attention outperforms the residual MLP for the same representation.}
\label{tab:top-configurations}
\setlength{\tabcolsep}{4pt}
\begin{tabularx}{\textwidth}{llXrcc}
\toprule
Split & Rank & Configuration & Runs & Accuracy & Macro-F1 \\
\midrule
Random & 1 & Morgan pair + Aux. reg. MLP & 5 & 0.905 $\pm$ 0.000 & \textbf{0.899 $\pm$ 0.000} \\
 & 2 & Morgan pair + MLP & 5 & 0.902 $\pm$ 0.001 & \underline{0.897 $\pm$ 0.001} \\
 & 3 & Morgan pair + Residual MLP & 5 & 0.902 $\pm$ 0.000 & 0.896 $\pm$ 0.001 \\
One drug unseen & 1 & GCN pair + Residual MLP & 5 & 0.672 $\pm$ 0.001 & \textbf{0.635 $\pm$ 0.001} \\
 & 2 & MoLFormer pair + Residual MLP & 5 & 0.669 $\pm$ 0.001 & \underline{0.632 $\pm$ 0.001} \\
 & 3 & Morgan pair + Residual MLP & 5 & 0.669 $\pm$ 0.003 & 0.631 $\pm$ 0.002 \\
Two drugs unseen & 1 & GCN pair + Residual MLP & 5 & 0.496 $\pm$ 0.005 & \textbf{0.432 $\pm$ 0.004} \\
 & 2 & GCN pair + MLP & 5 & 0.507 $\pm$ 0.004 & \underline{0.432 $\pm$ 0.007} \\
 & 3 & GCN pair + Aux. reg. MLP & 5 & 0.496 $\pm$ 0.002 & 0.426 $\pm$ 0.002 \\
Scaffold-disjoint & 1 & GCN pair + Pair attention & 5 & 0.343 $\pm$ 0.006 & \textbf{0.324 $\pm$ 0.005} \\
 & 2 & GCN pair + MLP & 5 & 0.345 $\pm$ 0.004 & \underline{0.319 $\pm$ 0.009} \\
 & 3 & GCN pair + Aux. reg. MLP & 5 & 0.336 $\pm$ 0.004 & 0.319 $\pm$ 0.004 \\
\bottomrule
\end{tabularx}
\end{table*}

%% file: rps_assets/tables/table_statistical_comparisons.tex
\begin{table*}[htbp]
\centering
\small
\caption{Paired five-seed macro-F1 differences for key benchmark comparisons. Differences are computed seed-wise as the first configuration minus the second. Intervals are 95\% paired $t$-intervals over seeds 11, 22, 33, 44, and 55.}
\label{tab:paired-comparisons}
\begin{tabularx}{\textwidth}{>{\raggedright\arraybackslash}p{0.25\textwidth}>{\raggedright\arraybackslash}X>{\raggedright\arraybackslash}p{0.16\textwidth}r}
\toprule
Comparison & Contrast & Split & Difference \\
\midrule
Random top configuration & Morgan aux. regression vs Morgan MLP & Random & 0.002 [0.001, 0.004] \\
One-drug representation margin & GCN residual vs MoLFormer residual & One drug unseen & 0.004 [0.002, 0.005] \\
Two-drug representation margin & GCN residual vs MoLFormer residual & Two drugs unseen & 0.012 [0.004, 0.020] \\
Scaffold representation margin & GCN pair attention vs MoLFormer residual & Scaffold-disjoint & 0.015 [0.009, 0.022] \\
Scaffold head margin & GCN pair attention vs GCN residual & Scaffold-disjoint & 0.009 [-0.002, 0.020] \\
\bottomrule
\end{tabularx}
\end{table*}

%% file: rps_assets/tables/table_probability_metrics.tex
\begin{table*}[htbp]
\centering
\scriptsize
\renewcommand{\arraystretch}{1.12}
\caption{Probability-based diagnostics for the leading five-seed configuration in each evaluated split. AUROC and AUPRC are macro-averaged one-vs-rest multiclass metrics recomputed from archived prediction probabilities. ECE is a 15-bin confidence calibration error.}
\label{tab:probability-diagnostics}
\resizebox{\textwidth}{!}{%
\begin{tabular}{llrrrr}
\toprule
Split & Configuration & Bal. acc. & MCC & AUROC & AUPRC / ECE \\
\midrule
Random & Morgan pair + auxiliary-regression MLP & 0.888 $\pm$ 0.001 & 0.875 $\pm$ 0.000 & 0.991 $\pm$ 0.000 & 0.964 $\pm$ 0.000 / 0.019 $\pm$ 0.001 \\
One drug unseen & GCN pair + residual MLP & 0.624 $\pm$ 0.003 & 0.563 $\pm$ 0.001 & 0.887 $\pm$ 0.001 & 0.670 $\pm$ 0.002 / 0.171 $\pm$ 0.003 \\
Two drugs unseen & GCN pair + residual MLP & 0.419 $\pm$ 0.003 & 0.330 $\pm$ 0.007 & 0.758 $\pm$ 0.001 & 0.415 $\pm$ 0.006 / 0.318 $\pm$ 0.003 \\
Scaffold-disjoint & GCN pair + pair attention & 0.325 $\pm$ 0.004 & 0.192 $\pm$ 0.007 & 0.689 $\pm$ 0.005 & 0.310 $\pm$ 0.005 / 0.322 $\pm$ 0.005 \\
\bottomrule
\end{tabular}%
}
\end{table*}

%% file: rps_assets/tables/table_split_difficulty_comparison.tex
\begin{table*}[htbp]
\centering
\small
\caption{Matched comparison between two-drug-unseen and scaffold-disjoint evaluation. Each row compares the same training setup, molecular representation, and model head across splits. Positive $\Delta$ indicates higher scaffold-disjoint macro-F1. Negative
$\Delta$ indicates higher two-drug-unseen macro-F1.}
\label{tab:split-comparison}
\begin{tabularx}{\textwidth}{Xrrrr}
\toprule
Matched group & Settings & Scaffold higher & Mean $\Delta$ & Median $\Delta$ \\
\midrule
All matched appendix settings & 36 & 0 & -0.085 & -0.093 \\
Torch settings & 28 & 0 & -0.093 & -0.098 \\
Morgan pair settings & 9 & 0 & -0.100 & -0.122 \\
GCN pair settings & 9 & 0 & -0.082 & -0.098 \\
Torch Morgan pair settings & 7 & 0 & -0.110 & -0.124 \\
Torch GCN pair settings & 7 & 0 & -0.091 & -0.106 \\
\bottomrule
\end{tabularx}
\end{table*}

%% file: rps_assets/tables/table_head_summary.tex
\begin{table*}[htbp]
\centering
\small
\caption{Classifier-head summary from the complete evaluated configuration set. Values average the five-seed means over the four pair representations within each split, so they are intended as supporting head-level evidence rather than a replacement for representation-specific comparisons.}
\label{tab:head-summary}
\begin{tabularx}{\textwidth}{llXcc}
\toprule
Split & Rank & Head & Mean accuracy & Mean macro-F1 \\
\midrule
Random & 1 & Residual MLP & 0.880 & 0.874 \\
 & 2 & Feature attention & 0.869 & 0.861 \\
 & 3 & Gated MLP & 0.868 & 0.860 \\
 & 4 & Aux. reg. MLP & 0.863 & 0.854 \\
One drug unseen & 1 & Residual MLP & 0.668 & 0.630 \\
 & 2 & Gated MLP & 0.663 & 0.619 \\
 & 3 & Feature attention & 0.660 & 0.616 \\
 & 4 & Aux. reg. MLP & 0.654 & 0.613 \\
Two drugs unseen & 1 & Residual MLP & 0.478 & 0.416 \\
 & 2 & Gated MLP & 0.475 & 0.401 \\
 & 3 & Aux. reg. MLP & 0.465 & 0.398 \\
 & 4 & MLP & 0.470 & 0.398 \\
Scaffold-disjoint & 1 & Residual MLP & 0.321 & 0.302 \\
 & 2 & MLP & 0.314 & 0.293 \\
 & 3 & Feature attention & 0.315 & 0.293 \\
 & 4 & Aux. reg. MLP & 0.313 & 0.293 \\
\bottomrule
\end{tabularx}
\end{table*}

%% file: rps_assets/bibliography.bib
@article{GilSorribes2026OvercomplexityDDI,
  title={Addressing model overcomplexity in drug-drug interaction prediction with molecular fingerprints},
  author={Gil-Sorribes, Manel and Molina, Alexis},
  journal={Journal of Cheminformatics},
  volume={18},
  number={1},
  pages={38},
  year={2026},
  doi={10.1186/s13321-025-01128-8},
  pmid={41709355},
  pmcid={PMC13014892}
}

@article{GilSorribes2026TensorDTI,
  title   = {{Tensor-DTI}: Enhancing Biomolecular Interaction Prediction
             with Contrastive Embedding Learning},
  author  = {Gil-Sorribes, Manel and Vilalta-Mor, J{\'u}lia and
             Filella-Merc{\`e}, Isaac and Soliva, Robert and
             Ciudad, {\'A}lvaro and Guallar, V{\'i}ctor and
             Molina, Alexis},
  journal = {arXiv preprint arXiv:2601.05792},
  year    = {2026},
  doi     = {10.48550/arXiv.2601.05792}
}

@article{Yu2021SumGNN,
  title={{SumGNN}: multi-typed drug interaction prediction via efficient knowledge graph summarization},
  author={Yu, Yue and Huang, Kexin and Zhang, Chao and Glass, Lucas M. and Sun, Jimeng and Xiao, Cao},
  journal={Bioinformatics},
  volume={37},
  number={18},
  pages={2988--2995},
  year={2021},
  doi={10.1093/bioinformatics/btab207}
}

@article{Knox2024DrugBank6,
  title={{DrugBank} 6.0: the {DrugBank} Knowledgebase for 2024},
  author={Knox, Craig and Wilson, Mike and Klinger, Christen M. and Franklin, Mark and Oler, Eponine and Wilson, Alex and Pon, Allison and Cox, Jordan and Chin, Na Eun and Strawbridge, Seth A. and others},
  journal={Nucleic Acids Research},
  volume={52},
  number={D1},
  pages={D1265--D1275},
  year={2024},
  doi={10.1093/nar/gkad976},
  pmid={37953279},
  pmcid={PMC10767804}
}

@article{Rogers2010ECFP,
  title={Extended-connectivity fingerprints},
  author={Rogers, David and Hahn, Mathew},
  journal={Journal of Chemical Information and Modeling},
  volume={50},
  number={5},
  pages={742--754},
  year={2010},
  doi={10.1021/ci100050t}
}

@article{Wu2018MoleculeNet,
  title={{MoleculeNet}: a benchmark for molecular machine learning},
  author={Wu, Zhenqin and Ramsundar, Bharath and Feinberg, Evan N. and Gomes, Joseph and Geniesse, Caleb and Pappu, Aneesh S. and Leswing, Karl and Pande, Vijay},
  journal={Chemical Science},
  volume={9},
  number={2},
  pages={513--530},
  year={2018},
  doi={10.1039/C7SC02664A}
}

@article{Bemis1996Murcko,
  title={The properties of known drugs. 1. Molecular frameworks},
  author={Bemis, Guy W. and Murcko, Mark A.},
  journal={Journal of Medicinal Chemistry},
  volume={39},
  number={15},
  pages={2887--2893},
  year={1996},
  doi={10.1021/jm9602928}
}

@article{Ross2022MoLFormer,
  title={Large-scale chemical language representations capture molecular structure and properties},
  author={Ross, Jerret and Belgodere, Brian and Chenthamarakshan, Vijil and Padhi, Inkit and Mroueh, Youssef and Das, Payel},
  journal={Nature Machine Intelligence},
  volume={4},
  pages={1256--1264},
  year={2022},
  doi={10.1038/s42256-022-00580-7}
}

@article{Chithrananda2020ChemBERTa,
  title   = {{ChemBERTa}: Large-Scale Self-Supervised Pretraining for
             Molecular Property Prediction},
  author  = {Chithrananda, Seyone and Grand, Gabriel and
             Ramsundar, Bharath},
  journal = {arXiv preprint arXiv:2010.09885},
  year    = {2020},
  doi     = {10.48550/arXiv.2010.09885}
}

@article{Gu2021PubMedBERT,
  title   = {Domain-Specific Language Model Pretraining for
             Biomedical Natural Language Processing},
  author  = {Gu, Yu and Tinn, Robert and Cheng, Hao and Lucas, Michael and
             Usuyama, Naoto and Liu, Xiaodong and Naumann, Tristan and
             Gao, Jianfeng and Poon, Hoifung},
  journal = {ACM Transactions on Computing for Healthcare},
  volume  = {3},
  number  = {1},
  pages   = {1--23},
  year    = {2021},
  doi     = {10.1145/3458754}
}

@article{Kipf2017GCN,
  title={Semi-supervised classification with graph convolutional networks},
  author={Kipf, Thomas N. and Welling, Max},
  journal={International Conference on Learning Representations},
  year={2017},
  url={https://openreview.net/forum?id=SJU4ayYgl}
}

@article{Kim2023PubChem,
  title={{PubChem} 2023 update},
  author={Kim, Sunghwan and Chen, Jie and Cheng, Tiejun and Gindulyte, Asta and He, Jia and He, Siqian and Li, Qingliang and Shoemaker, Benjamin A. and Thiessen, Paul A. and Yu, Bo and Zaslavsky, Leonid and Zhang, Jian and Bolton, Evan E.},
  journal={Nucleic Acids Research},
  volume={51},
  number={D1},
  pages={D1373--D1380},
  year={2023},
  doi={10.1093/nar/gkac956},
  pmid={36305812},
  pmcid={PMC9825602}
}

@misc{Landrum2025RDKit,
  author       = {Landrum, Greg and Tosco, Paolo and Kelley, Brian and others},
  title        = {{RDKit}: Release 2024\_09\_5},
  howpublished = {Zenodo software release},
  year         = {2025},
  doi = {10.5281/zenodo.14779836}
}

@inproceedings{Gilmer2017NeuralMessagePassing,
  title     = {Neural Message Passing for Quantum Chemistry},
  author    = {Gilmer, Justin and Schoenholz, Samuel S. and
               Riley, Patrick F. and Vinyals, Oriol and Dahl, George E.},
  booktitle = {Proceedings of the 34th International Conference on
               Machine Learning},
  volume    = {70},
  pages     = {1263--1272},
  publisher = {PMLR},
  year      = {2017}
}

@inproceedings{Reimers2019SentenceBERT,
  author    = {Reimers, Nils and Gurevych, Iryna},
  title     = {{Sentence-BERT}: Sentence Embeddings Using Siamese
               {BERT}-Networks},
  booktitle = {Proceedings of the 2019 Conference on Empirical Methods
               in Natural Language Processing and the 9th International
               Joint Conference on Natural Language Processing},
  pages     = {3982--3992},
  publisher = {Association for Computational Linguistics},
  address   = {Hong Kong, China},
  year      = {2019},
  doi       = {10.18653/v1/D19-1410}
}

@inproceedings{Sculley2010WebScaleKMeans,
  author    = {Sculley, D.},
  title     = {Web-Scale K-Means Clustering},
  booktitle = {Proceedings of the 19th International Conference on World Wide Web},
  series    = {WWW '10},
  pages     = {1177--1178},
  publisher = {Association for Computing Machinery},
  address   = {New York, NY, USA},
  year      = {2010},
  doi       = {10.1145/1772690.1772862}
}

@article{Zhong2024MeTDDI,
  author = {Yi Zhong and Gaozheng Li and Ji Yang and Houbing Zheng and Yongqiang Yu and Jiheng Zhang and Heng Luo and Biao Wang and Zuquan Weng},
  title = {Learning motif-based graphs for drug-drug interaction prediction via local-global self-attention},
  journal = {Nature Machine Intelligence},
  volume = {6},
  pages = {1094--1105},
  year = {2024},
  doi = {10.1038/s42256-024-00888-6}
}

@inproceedings{Guo2017Calibration,
  title     = {On Calibration of Modern Neural Networks},
  author    = {Guo, Chuan and Pleiss, Geoff and Sun, Yu and
               Weinberger, Kilian Q.},
  booktitle = {Proceedings of the 34th International Conference on
               Machine Learning},
  volume    = {70},
  pages     = {1321--1330},
  publisher = {PMLR},
  year      = {2017}
}
